\documentstyle[12pt]{article} 
\begin{document}

\newpage

{\bf     On Superluminal motions in photon and particle tunnellings.}

\vspace{7mm}

\hspace{5mm}     J.Jakiel, {\em Institute of Nuclear Physics, 
31-342 Krak\'{o}w, Poland.}

\hspace{5mm}  V.S.Olkhovsky, {\em Institute for Nuclear Research, 252028
Kiev, Ukraine.}

\hspace{5mm}   E.Recami, {\em Facolta' di Ingegneria, Universita' statale di
Bergamo,

\hspace{5mm}     24044 Dalmine (BG), Italy; I.N.F.N--Sezione di Milano, Milan,
Italy,}

\hspace{5mm}  and {\em D.M.O./FEEC and C.C.S., UNICAMP, Campinas, S.P., Brazil.}

\vspace{15mm}

Abstract: It is shown that the Hartman--Fletcher effect is valid
for all known expressions of the mean tunnelling time, in various
nonrelativistic approaches, for the case of finite width barriers
without absorption. Then, we show that the same
effect is not valid for the tunnelling time mean-square fluctuations. 
On the basis of the Hartman--Fletcher effect and the known analogy between
photon and nonrelativistic-particle tunnelling, one can
explain the Superluminal group-velocities observed in
various photon tunnelling experiments (without violation of the so-called
``Einstein causality").

\vspace{7mm}

\newpage

\qquad 1. The Hartman--Fletcher effect (HFE) was first
studied in~[1,2] within the stationary-phase method for
quasi-monochromatic nonrelativistic particles tunnelling through
potential barriers. It consists in the result that
the tunnelling {\em phase-time}
\begin{equation}
  \tau_{\rm tun}^{\rm Ph}=\hbar (\partial (\arg A_{\rm T}+ka))/(\partial E)
\end{equation}
is independent of the barrier width $a$ for sufficiently large $a$; \
where $\tau^{\rm Ph}_{\rm tun}$ is {\em the mean tunnelling time} 
$ \langle \tau_{\rm tun} \rangle $ evaluated by the stationary-phase method,
when it is possible to neglect the interference between incident and
reflected waves before the barrier;  
while $A_{\rm T}$ and
$E=\hbar^{2}k^{2}/(2\mu)$  are transmission amplitude and
particle kinetic energy, respectively. \
In particular, for a rectangular potential
barrier  $\tau_{\rm tun}^{\rm Ph} \rightarrow 2/(\upsilon \kappa)$ when
$\kappa a >>1 $, quantity $\upsilon =\hbar \kappa/\mu$ being the particle
velocity. \ \ It is known that, within different approaches and
under various conditions, many different theoretical expressions for the
mean tunnelling time were deduced in the literature (see, for instance,~[3]).
The first goal of this paper is to analyze the role and validity of the HFE
for all its known theoretical expressions. Our second
goal consists in explaining the recent photon tunnelling experiments,
in which Superluminal phenomena were observed, on the basis of the mentioned
HFE and the analogy between photon and  nonrelativistic-particle tunnelling.\\

\qquad 2. If one chooses, out of all the known definitions of tunnelling
times, the {\em mean dwell time} $ \langle \tau_{\rm tun}^{\rm dw} 
\rangle $~[4], \ the {\em first mean Larmor time} $ \langle 
\tau_{\rm tun}^{\rm L} \rangle $~[4,5], \  and the real part
of the complex tunnelling time obtained {\em by averaging over the
Feynman paths} ${\rm Re} \; \tau_{\rm tun}^{\rm F}$~[6], which equal  $\hbar
k/(\kappa V_{0})$  for quasi-monochromatic particles and opaque
rectangular barriers, one immediately and easily sees that also in such
cases there is {\em no} dependence on the barrier width, and
consequently the HFE is present.

\qquad The validity of the HFE for the {\em mean tunnelling time}, within
the general nonrelativistic approach developed in~[3,7-11], can be
directly inferred from the expression
\begin{equation}
 \langle \tau_{\rm tun} \rangle  =
 \langle t_{+}(a) \rangle  -  \langle t_{+}(0) \rangle   =   \langle 
 \tau^{\rm Ph}_{\rm tun} \rangle _{E} -  \langle t_{+}(0) \rangle 
\end{equation}
that was found in~[8-11], and was additionally confirmed in~[7-11] by
numerous calculations for various cases of gaussian electron wavepackets.
In eq.(2), the symbol $ \langle ... \rangle _{E}$ denotes the average over 
the initial wavepacket energy spread, and it is
 $ \langle t_{+}(x) \rangle  =\int^{\infty}_{-\infty} tJ_{\pm}(x,t)dt /
 \int^{\infty}_{-\infty} J(x,t)dt$,
with $J_{\pm}(x,t) = \theta (\pm J)\cdot J(x,t)$, where $J(x,t)$ is
the probability flux density for a wavepacket moving
through a barrier located in the interval $(0,a)$ along the $x$-axis. \
Since our
conclusion about the validity of the HFE is true for rectangular barriers,
one can easily be convinced that it holds also for more general
barriers, localized in the interval $(0,a)$.

\qquad Let us stress that up to now {\em there is no a unique general
formulation of the causality condition, necessary and sufficient for
all possible cases of collisions}. Requiring non-negative values for the mean
durations represent the most simple causality condition (which is
sufficient, but not necessary). Although in the approach presented
in~[3,7-11] there is no {\em direct} causal connection between the
peaks of the incoming  and of the transmitted or
reflected wavepackets (because of the nontrival
motions inside the barrier), the previous causality condition seems to be
valid for the whole tunnelling[10,11]. An analysis of various
formulations of the causality condition is presented in~[10,11] for both
nonrelativistic and relativistic tunnelling.

\qquad In any case, concluding this section, we may assert that no
exception is known of the validity of the HFE, for whatever expression of
the mean tunnelling time.\\

\qquad 3. Let us recall that the second Larmor time[4]
\begin{equation}
\tau^{\rm L}_{z,{\rm tun}}= \hbar [ \langle (\partial \mid A_{\rm T} \mid / \partial
E)^{2} \rangle / \langle  \mid A_{\rm T} \mid ^2 \rangle ]^{1/2}
\end{equation}
as well as the {\em B\"{u}ttiker-Landauer time} \ $\tau_{\rm tun}^{\rm B-L}$,[12] \
and the {\em imaginary part of the complex tunnelling time} \ ${\rm Im} \;
\tau_{\rm tun}^{\rm F}$ (obtained within the Feynman approach[6]) ---both of
which are equivalent to expression (3)---, in the limiting case of opaque
rectangular barriers yield all the same value $a \mu/(\hbar \kappa)$, i.e.
become all proportional to the barrier width  $a$. It was shown in~[10,11],
however, that those times are {\em not}
mean times, but {\em mean-square fluctuations} of the tunnelling-time
distribution. In fact, they are equal to \ $[D_{\rm dyn} \tau_{\rm tun}]^{1/2}$, \
where $D_{\rm dyn} \tau_{\rm tun}$ is the tunnelling-time dynamical variance
caused by the barrier only and defined by the equation \
$D_{\rm dyn} \; \tau_{\rm tun} = D\tau_{\rm tun} - Dt_{+}(0)$, \ with
$D \tau_{\rm tun} =  \langle \tau_{\rm tun}^{2} \rangle  -  \langle 
\tau_{\rm tun} \rangle ^{2}$; \ and \ $ \langle \tau_{\rm tun}^{2} \rangle  =  
\langle [t_{+}(a) - \langle t_{+}(0) \rangle ]^{2} \rangle  + Dt_{+}(0)$. \ \
Hence they are {\em not} connected with the peak (or
group) velocities of tunnelling particles, {\em but} rather with the relevant
tunnelling velocity {\em distribution} along the barrier region.\\

\qquad 4. All these results have been obtained for transparent media
(without absorption and/or dissipation). As it was theoretically
demonstrated in~[13] within nonrelativistic quantum mechanics, {\em the
HFE vanishes for barriers with strong enough absorption}. This was confirmed
experimentally in the case of electromagnetic (microwave) tunnelling in~[14].
Namely, it follows from ref.[13] that, if one describes the absorption by
adding the imaginary term  $-iV_{1} \ \; (V_{1} >0)$ to $V_{0}$, then the
HFE does not disappear only for very small absorptions, when $V_{1} << V_{0}$ 
and \ $V_{1} \mu^{1/2} \nu \kappa a /[2(V_{0} -E)]^{3/2} << 2$.\\

\qquad 5. Let us now propose, here, another description of tunnelling which
appears convenient for applications to media without absorption and
dissipation, and for Josephson junctions. \ In our new
representation the transmission and reflection amplitudes are
rewritten (for the same external boundary conditions[15]) as
\begin{equation}
    A_{\rm T} = i \, {\rm Im} \, (\exp (i \varphi_{1})) \exp (i \varphi_{2}-ika); \
    A_{\rm R} =  {\rm Re} \, (\exp (i \varphi_{1})) \exp (i \varphi_{2}-ika)
\end{equation}
where for rectangular potential barriers it is \
$\varphi_{1} = \arctan\{2 \sigma /[(1+ \sigma^{2}) \sinh (\kappa a)]\}$; \ and \
$\varphi_{2} = \arctan\{ \sigma \sinh(\kappa a)/[\sinh^{2} (\kappa
a/2) -\sigma^{2} \cosh^{2} ( \kappa a/2)]\}$, \
with \ $\sigma = \kappa /k $, and $\kappa^{2} = \kappa^{2}_{0} - k^{2}$, \
quantity $\kappa_{0} =[2\mu V_{0}]^{1/2} / \hbar$ being  the barrier height. \
We have introduced two new phases, $\varphi_{1}$ and $\varphi_{2}$, in terms
of which the expressions for $\tau_{\rm tun}^{\rm Ph}$ and for
$\tau_{z,{\rm tun}}^{\rm L} = \tau_{\rm tun}^{\rm B-L}$ acquire
the following form:

\begin{equation}
  \tau_{\rm tun}^{\rm Ph}=\hbar \frac{\partial (\arg A_{\rm T})}{\partial E}=\hbar
  \frac{\partial (\varphi_{2}-ka)}{\partial E}, \;
  \tau_{z,{\rm tun}}^{\rm L}=\tau_{\rm tun}^{\rm B-L}=\hbar \frac{\partial
  \varphi_{1}}{\partial E} \cot(\varphi_{1}) \ .
\end{equation}

So, we see that, in the rectangular opaque potential limit, without
absorption, the phase $\varphi_{2} -ka$ leads to the HFE for tunnelling
times, while the phase $\varphi_{1}$ leads to a dependence of
tunnelling times on $a$. \ For the times $ \langle \tau_{\rm tun}^{\rm dw}  \rangle  =
 \langle \tau_{y,{\rm tun}}^{\rm L} \rangle $ we obtain a complicated formula, which can be
expressed in terms of $\varphi_{2}$ and of $\partial \varphi_{2}/\partial E$
only at the limit $\kappa a >>1$. \ In the presence of absorption, both phases
become complex and expressions (5) are much bulkier and, in general, depend
on $a$ (with a violation of the HFE, in accordance with refs.[13,14]).\\

\qquad 6. These conclusions can be easily extended for photon
tunnelling, as already mentioned, if one takes account of the analogy
between the photon and nonrelativistic-particle tunnelling. Such an analogy
was first examined in refs.[16,17] for the case of the time-independent
Schr\"{o}dinger and Helmholtz equations, and in refs.[10,11] for the case
of the time-dependent Schr\"{o}dinger and Maxwell equations (on the basis,
also, of a suitable generalization of our operator for Time from
quantum mechanics to first-quantization quantum electrodynamics).

\qquad While for nonrelativistic particle tunnelling is associated with
potentizl barriers, in the photon experiments tunnelling takes place
through different kinds of regions generating
evanescent (decreasing) and anti-evanescent (increasing) waves: for instance,
waveguides with cutoff (undersized) regions[18,19], multilayer dielectric
mirrors as realizations of one-dimensional (1D) photonic band gaps[20,21],
and frustrated total internal reflection regions[22]. \
In all these cases the HFE can bring to so large group (or peak, or effective)
velocities of the tunnelling photons that they can acquire {\em
Superluminal\/}[18-22] and even infinite[23], or negative[24], speed.

\qquad Here we shall briefly analyze the Superluminal phenomena
observed in the 1D microwave tunnelling experiment and 2D optical
tunnelling experiment described in~[14] and~[22], respectively.  

\qquad  In both cases, inside the ``photonic barrier" (similarly to what done
for the quantum-mechanical wave functions) the electromagnetic waves 
are to be described by the superposition of evanescent and
anti-evanescent waves: \
 $\alpha \exp (-\kappa_{\rm e} x) + \beta \exp (\kappa_{\rm e} x)$, \   with
{\em non-zero} fluxes for non-zero complex coefficients $\alpha$ and $\beta$,
which depend on the boundary conditions. The evanescent-wave wave number
$\kappa_{\rm e}$  is defined by the difference between the incident wave
frequency and the cutoff frequency in the microwave experiment[14,18]: \
$\kappa_{\rm e}=2\pi[\lambda^{-2} -\lambda_{\rm c}^{-2}]^{1/2}$, \ quantity 
$\lambda_{\rm c}$ being the cutoff wave length, {\em and} by the difference
between the incidence angle  $\hat{\imath}$ and the critical angle  
${\hat{\imath}}_{\rm c} = \sin^{-1}(1/n)$ of total reflection 
(${\hat{\imath}} > {\hat{\imath}}_{\rm c}$; \ $n>1$) in the optical 
experiment[22]. \
For quasi-monochromatic wavepackets and appropriate
measurement conditions (which were realized in the experiments
described in~[18,20-22]), one can use the stationary-phase method,  and
the phase tunnelling time is defined by the expression[10,11]
\begin{equation}
		\tau_{\rm tun}^{\rm Ph}= \frac{2}{c\kappa}
\end{equation}
for  $\kappa a >>1 $. \ \ The quantity \ $\upsilon_{\rm tun}^{\rm eff} =
a/\tau_{\rm tun}^{\rm Ph}$, \ which has the meaning of the effective photon
tunnelling velocity, resulted in both experiments~[18,22] (and also in~[20,21])
to be Superluminal, since $\kappa_{\rm e} > 2$. \  In the microwave experiment[18]
the time $\tau_{\rm tun}^{\rm Ph}$  was found by pulsed measurements in the
time domain, and expression (6) was directly tested[10,11]. \  In the optical
experiment[22], the quantity $\tau_{\rm tun}^{\rm Ph}$  was calculated with
the help of the simple formula
\begin{equation}
       \tau_{\rm tun}^{\rm Ph} = \frac{D}{\upsilon_{x}} \; ; \qquad
	\upsilon_{x} = c/(n \sin({\hat{\imath}}))
\end{equation}
by measuring $D$  and  $\hat{\imath}$.

\qquad  It was also shown in~[22] that the angular deviation  $\delta
{\hat{\imath}}$  of the emerging beam is related to the ``loss time"
$\tau_{\rm tun}^{\rm loss}$ , which is precisely equal to expression (2),
and so has the meaning of a {\em tunnelling time mean-square fluctuation}, by
the formula
\begin{equation}
		    \delta {\hat{\imath}} = \Omega \; \tau_{\rm tun}^{\rm loss} \ ,
\end{equation}
where the frequency $\Omega$ is determined by  $\hat{\imath}$, $n$  and  the beam
Rayleigh length . Let us remind again that the quantity $\tau_{\rm tun}^{\rm loss}$
is connected not with the effective photon tunnelling velocity, but with
the {\em tunnelling-velocity distribution}.\\

\qquad 7. Thus, we have seen that {\em the HFE is valid for all the known
expressions of the mean tunnelling times 
(without absorption and dissipation)}, while it is not valid for the
tunnelling time mean-square fluctuations. {\em Then, the HFE is a good basis 
for explaining the Superluminal group velocities of tunnelling photons}. \
These are our main conclusions, substantiated in Sects.2--6.
	 
\qquad  Finally, let us mention that many discussions (see e.g. refs.[25-28])
were generated by the Superluminal phenomena, observed e.g. in the
experiments~[18-22], and revived by the results observed in similar
phenomena, namely the electromagnetic pulse propagation in dispersive 
media[29-31]. \ In fact, on the other hand, it is known since long time that 
the {\em wave-front velocity}
for electromagnetic pulse propagation cannot exceed the vacuum light
speed $c$.~[32,33] \ A conclusion confirmed by various methods and in 
various processes, including tunnelling,[25-28] and that seems to guarantee 
the so-called (naive) ``Einstein causality". \ We shall come
back to these points[23] elsewhere.\\

\qquad {\bf Acknowledgments} --  This work was partially supported by INFN, 
MURST and CNR. \ The authors thank
A.Budzanowski, G.Degli Antoni, F.Fontana, H.E.Hern\'andez F.,  L.C.Kretly 
and J.W.Swart for scientific help and stimulating discussions, and
A.Agresti, Y.Akebo, A.Pablo L.Barbero, R.Bonifacio, A.Gabovich,
R.Garavaglia, A.Shaarawi, A.Zaichenko and M.Zamboni-Rached for kind scientific
collaboration.

\newpage

R e f e r e n c e s \\

1. T.E.Hartman, J. Appl. Phys. 33, 3427 (1962).

2. J.R.Fletcher, J. Phys. C18, L55 (1985).

3. V.S.Olkhovsky and E.Recami, Phys. Rep. 214, 339 (1992).

4. M.Buettiker, Phys. Rev. B27, 6178 (1983).

5. A.Baz', Sov. J. Nucl. Phys. 4, 182 (1967); 5, 161 (1967).             
								
6. D.Sokolovski and L.Baskin, Phys. Rev. A36, 4604 (1987).       

7. V.S.Olkhovsky, E.Recami and A.K.Zaichenko, Solid State Commun.
89, 31 (1994).

8. V.S.Olkhovsky, E.Recami, F.Raciti and A.K.Zaichenko, J. de Phys.-I
(France) 5, 1351 (1995).

9. V.S.Olkhovsky and A.K.Zaichenko, Ukrain. Fiz. Zhurnal  42, 751
(1997).

10. V.S.Olkhovsky and A.Agresti, in {\em Tunnelling and its Implications} (World
Scient.; Singapore, 1997), pp.327-355.

11. V.S.Olkhovsky, Physics of the Alive  5, 23 (1997).

12. M.Buettiker and R.Landauer, Phys. Rev. Lett. 49, 1739 (1982);
Phys. Scr. 32, 429 (1985).

13.  F.Raciti and G.Salesi, J. de Phys.-I (France) 4, 1783 (1994).

14. G.Nimtz, H.Spieker and M.Brodowsky, J. de Phys.-I (France) 4, 1379
(1994).

15. A.Messiah, {\em Quantum Mechanics} (North-Holland; Amsterdam, 1961)
(Vol.1, Chapt.3.11).

16. R.Y.Chiao, P.G.Kwiat and A.M.Steinberg, Physica  B175, 257 (1991).

17. Th.Martin and R.Landauer, Phys. Rev. A45, 2611 (1992).

18. A.Enders and G.Nimtz, J. de Phys.-I (France) 2, 1693 (1992); 3, 1089
(1993); Phys. Rev. B47, 9605 (1993); Phys. Rev. E48, 632 (1993); \
H.M.Brodowsky, W.Heitmann and G.Nimtz: Phys. Lett. A222, 125 (1996).

19. A.Ranfagni, P.Fabeni, G.P.Pazzi and D.Mugnai, Phys. Rev. E48, 1453
(1993).

20. A.M.Steinberg, P.G.Kwiat and R.Y.Chiao, Phys. Rev. Lett. 71
(1993) 708; \ R.Y.Chiao, P.G.Kwiat and A.M.Steinberg: Scientific American 
269 (1993), issue no.2, p.38. \ Cf. also P.G.Kwiat et al.: Phys. Rev. A48 
(1993) R867; \ E.L.Bolda et al.: Phys. Rev. A48 (1993) 3890; \ A.M.Steinberg 
and R.Y.Chiao: Phys. Rev. A51 (1995) 352. 

21. Ch.Spielman, R.Szip"cs, A.Stingl and F.Krausz, Phys. Rev. Lett. 73,
2308 (1994).

22. Ph.Balcou and L.Dutriaux, Phys. Rev. Lett. 78, 851 (1997).

23. A.P.L.Barbero, H.E.Hern\'andez F., and E.Recami, ``On the propagation
speed of evanescent modes", submitted for pub.

24. E.Recami, Rivista Nuovo Cim. 9 (1986), issue no.6, pp.1-178, and refs.
therein. \ See also refs.[29,31].

25. J.M.Deutch and F.E.Low, Ann. of Phys. 228, 184 (1993); \ K.Hass and
P.Busch, Phys. Lett. A185, 9 (1994); \ M.Ya.Azbel', Solid State Commun. 91,
439 (1994).

26. W. Heitman and G.Nimtz, Phys. Lett. A196, 154 (1994).

27. G.Nimtz, A.Enders and H.Spieker, J. de Phys.-I (France) 4, 565 (1994).

28. A.Ranfagni and M.Mugnai, Phys. Rev. E52, 1128 (1995);  D.Mugnai,
A.Ranfagni,
 \ \ \      R.Ruggeri, A.Agresti and E.Recami, Phys. Lett. A209, 227 (1995).

29.  C.G.B.Garrett and D.E.McCumber, Phys. Rev. A1, 305 (1970).

30. M.D.Crisp, Phys. Rev.  A4, 2104 (1971).

31. S.Chu and S.Wong, Phys. Rev. Lett. 48, 738 (1982). \ Cf. also R.Y.Chiao,
A.E.Kozhekin and G.Kurizki: Phys. Rev. Lett. 77, 1254 (1996); \ M.W.Mitchell
and R.Y.Chiao: Phys. Lett. A230, 133 (1997).

32. A.Sommerfeld, Z. Phys. 8, 841 (1907); 44, 177 (1914).

33. L.Brillouin, {\em Wave Propagation and Group Velocity} (Acad. Press;
New York, 1960).

\end{document}